\documentclass[12pt]{iopart} 
\expandafter\let\csname equation*\endcsname\relax 

\expandafter\let\csname endequation*\endcsname\relax 
\usepackage{amsmath} 
\usepackage{multirow}
\usepackage{mathtools}
\usepackage{xparse}
\usepackage[super]{nth}

\newcommand{\ket}[1]{\left\vert#1\right\rangle}
\newcommand{\bra}[1]{\left\langle#1\right\vert}
\newcommand{\comm}[2]{\left[#1,#2\right]}

\newcommand{\average}[1]{\left\langle#1\right\rangle}

\DeclarePairedDelimiter\ceil{\lceil}{\rceil}
\DeclarePairedDelimiter\floor{\lfloor}{\rfloor}
 
\begin{document}

\title{Perturbatively-perfect many-body transfer}

\author{Wayne Jordan Chetcuti\dag, Claudio Sanavio\dag, Salvatore Lorenzo\ddag, Tony J. G. Apollaro\dag \footnote{To
		whom correspondence should be addressed.}}

\address{\dag\ Department of Physics, Faculty of Science, University of Malta, Msida MSD 2080, Malta}
\address{\ddag\ Dipartimento di Fisica e Chimica - Emilio Segr\`{e}, Universit\`{a} degli Studi di Palermo, via Archirafi 36, I-90123 Palermo, Italy}

\ead{tony.apollaro@um.edu.mt}
\vspace{10pt}

\begin{abstract}
The coherent transfer of excitations between different locations of a quantum many-body system is of primary importance in many research areas, from transport properties in spintronics and atomtronics to quantum state transfer in quantum information processing. We address the transfer of $n>1$ bosonic and fermionic excitations between the edges of a one-dimensional chain modeled by a quadratic hopping Hamiltonian, where the block edges, embodying the sender and the receiver sites, are weakly coupled to the quantum wire. We find that perturbatively perfect coherent transfer is attainable in the weak-coupling limit, for both bosons and fermions, only for certain modular arithmetic equivalence classes of the wire's length. Finally we apply our findings to the transport of spins and the charging of a many-body quantum battery.
\end{abstract}

%
%
%
%
%

\section{Introduction}
Quantum many-body dynamics lies at the core of most of the theoretical and experimental physics~\cite{bruus2004many}. Applications of quantum many-body dynamics are found in countless technologies, ranging from electronics to spintronics where characterising transport properties is of paramount importance~\cite{Amico_2017, RevModPhys.76.323}.
However, quantum many-body systems are notoriously difficult to solve. Already finding the ground state of a one-dimensional two-body local Hamiltonian is a QMA-complete problem~\cite{2004quant.ph..6180K} - let alone its dynamics- and many strategies have been proposed to tackle the many-body problem, from DMRG to Quantum Montecarlo, just to name a few algorithms, as well as quantum simulators~\cite{Bernien2017}. A notable exception is constituted by the class of integrable models~\cite{Franchini:2016cxs}, where analytical methods are available for determining the full spectrum of the Hamiltonian. Still, a complete characterisation of the dynamical behaviour of, say, an observable is a formidable task. 
The class of quadratic Hamiltonians in creation and destruction bosonic and fermionic operators embody a small, but significant, subset of integrable models. Their computationally manageable dynamics rests on the fact that they can be mapped onto non-interacting models.

Recently, significant steps forward have been  achieved experimentally in simulating many-body systems, e.g., with cold atoms~\cite{Bloch2012}. Likewise, the capacity of manipulating single- or few-body subsets of a many-body system~\cite{Weitenberg2011} is becoming key for many quantum devices, spanning from quantum information to quantum thermodynamics applications. In these experiments, significant attention has been devoted to the coherent transport of excitations along one-dimensional quantum systems~\cite{Ramanathan_2011, Hirobe2016, Fukuhara2013, PhysRevX.7.041063}.
 
The transfer of excitations between edges of a spin chain, some instances of which can be mapped to a quadratic Hamiltonian, has been addressed in several works, with particular emphasis given to the quantum state transfer of a single qubit in quantum information processing. Fully engineered wires are able to achieve such a goal with unit fidelity in a ballistic time~\cite{PhysRevLett.92.187902,Kaur_2012, PhysRevLett.101.230502, Chapman2016}. Nevertheless, a precise control over each coupling constant is experimentally demanding, especially in solid state systems. Alternative methods have been proposed where only a few couplings are required to be addressed, generally being that between the sender (receiver) site and the quantum channel ~\cite{qst1, PhysRevA.93.032310,Banchi_2011, PhysRevA.85.052319, PhysRevA.98.012334, VIEIRA20182586}. 
The case of an higher number of excitations, or the transfer of an
arbitrary two-qubit state, has received less attention~\cite{1367-2630-16-12-123003, doi:10.1142/S021974991750037X, doi:10.1142/S021974991750037X, 1402-4896-2015-T165-014036}, whereas the transfer of a state of more than two qubits has not been addressed yet in a setting where the quantum channel is made up of a chain with uniform couplings.  

In this paper we address the problem of the coherent transfer of $n>1$ excitations between the edges of a system described by a 1D quadratic many-body Hamiltonian. Due to the Hamiltonian's non-interacting nature, we are able to express the many-body dynamics in terms of one-body transition amplitudes. Exploiting this property, we identify the equivalence classes for the length of the 1D system for which the coherent transfer for up to four excitations occurs, regardless of their bosonic or fermionic nature. The transfer happens, for specific lengths of the chain, via Rabi-like dynamics in the weak-coupling regime, which we consequently dub as PP (perturbatively perfect)  excitation transfer. 

The paper is organised as follows. In Sec.~\ref{S.Model} the nearest-neighbor hopping Hamiltonian, the setup for $n$ excitation transfer and its transition amplitude matrix are introduced. Moreover, the definition of PP transfer and the single-particle eigenenergies resonance conditions between the sender (receiver) block and the wire energy spectrum are defined. In Sec.~\ref{Ss.MBD1} the many-body dynamics for up to four excitations in the sender block is analysed for each of the equivalence classes of the wire defined by the resonance conditions. In Sec.~\ref{S.Applications}, the $n$ excitations dynamics is applied to magnetisation and energy transport. Finally, in Sec.~\ref{S.Concl} we draw our conclusions.

\section{The model}\label{S.Model}
\begin{figure}[h!]
	\label{F.Figure}
	\includegraphics[width =\textwidth]{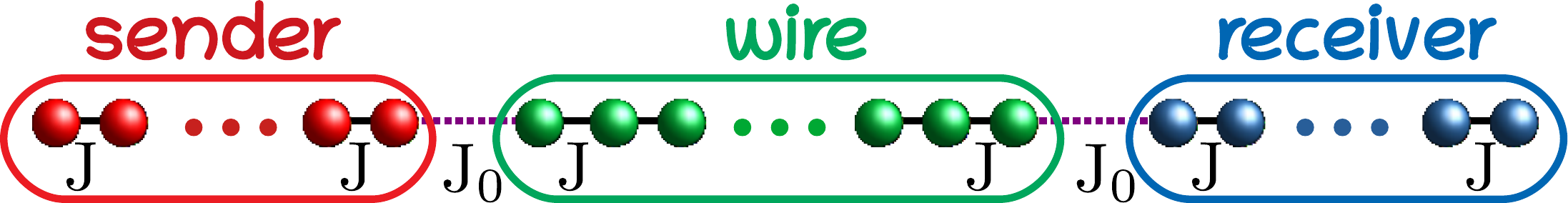}
	\caption{Setup of the excitation transfer protocol. Sender and receiver block, with the excitations residing in the former, are weakly coupled by $J_0$ at both edges of a wire. Each part is made up by a 1D lattice described by the Hamiltonian in Eq.~\ref{E_Ham1} with $J_i=J=1$, but for $J_0\ll 1$, and $h_i=h$.}
\end{figure}

We consider a hopping Hamiltonian with nearest-neighbor interaction $J_i$ and an on-site potential $h_i$ on a 1D lattice 
\begin{align}
\label{E_Ham1}
\hat{H} = \sum\limits_{i=1}^{N}\frac{J_i}{2}\left(\hat{c}_{i+1}^{\dagger}\hat{c}_{i}+\hat{c}_{i}^{\dagger}c_{i+1}\right)+ h_i \hat{c}_i^{\dagger}\hat{c}_i~,
\end{align}
where the $\hat{c}$'s represent either fermions or bosons, and open boundary conditions are assumed, $\hat{c}_{N+1}=\hat{c}^{\dagger}_{N+1}=0$. 
In the subsequent sections, we will assume that the couplings $J_i$ are all uniform but for the couplings $J_i=J_0$ between the sender (receiver) block and the wire (see Fig.~\ref{F.Figure}). We will also set the coupling within the sender (receiver) block and within the wire as our time and energy unit $J_i=J=1$. In the present section, these assumptions are unnecessary for the diagonalisation of the model we are going to outline.

As the number operator, $\hat{\mathcal{N}}=\sum_{i=1}^N \hat{c}_i^{\dagger}\hat{c}_i$, commutes with the Hamiltonian in Eq.~\ref{E_Ham1} as a consequence of the $U(1)$ symmetry of the model, the total number of excitations is conserved. To show this symmetry explicitly, we express the Hamiltonian as a direct sum of invariant-subspace Hamiltonians, $\hat{H}=\bigoplus_{i=0}^N \hat{H}^{\left(n\right)}$, where $n$ denotes the number of excitations. Due to the non-interacting, i.e., quadratic, nature of the Hamiltonian, single-particle eigenstates are sufficient to investigate the full many-body dynamics.

The hopping Hamiltonian in Eq.~\ref{E_Ham1} in the single-particle sector is easily diagonalised as
\begin{align}\label{E_1part}
\hat{H}^{\left(1\right)}=\sum_{k=1}^N \omega_k \ket{\phi_k}\!\!\bra{\phi_k}\equiv \sum_{k=1}^N \omega_k \hat{c}_k^{\dagger}\hat{c}_k~,
\end{align}
where $\{\omega_k,\ket{\phi_k}\equiv \hat{c}_k^{\dagger}\ket{0}\}$ are the eigenvalues and the eigenvectors of the tridiagonal matrix, $A\equiv \bra{i}\hat{H}^{\left(1\right)}\ket{j}=\frac{J_i}{2}\left(\delta_{i,i+1}+\delta_{i,i-1}\right)+h_i \delta_{i,j}$, describing the single-particle dynamics in the direct space basis, $\ket{i}\equiv \hat{c}_i\ket{0}$. Here, and in the following, $\ket{i}\equiv\ket{00\dots 1_i\dots 00}$ represents a state with one excitation sitting on site $i$.  The symbol $A$ has been used to stress the equivalence between $\hat{H}^{\left(1\right)}$ and the adjacency matrix used in graph theory~\cite{PhysRevLett.92.187902}.

From Eq.~\ref{E_1part} it is straightforward to explicit
the spectral decomposition of each Hamiltonian $\hat{H}^{\left(n\right)}$ in the direct-sum expansion, which for fermionic excitations reads
\begin{align}
\label{E.Spectralf}
\hat{H}^{\left(n\right)}=\sum_{p_1<p_2<\dots<p_n=1}^N\left(\omega_{p_1}+\omega_{p_2}+\dots+\omega_{p_n}\right)\hat{c}_{p_n}^{\dagger}\dots\hat{c}_{p_2}^{\dagger}\hat{c}_{p_1}^{\dagger}\hat{c}_{p_1}\hat{c}_{p_2}\dots\hat{c}_{p_n}~.
\end{align}
Clearly the full fermionic $2^N$-dimensional Hamiltonian is retrieved by summing all the binomial-dimensional $\hat{H}^{\left(n\right)}$, as $\sum_{n=0}^{N}\binom{N}{n}=2^N$.
For bosonic excitations, on the other hand, the system is described by an infinite-dimensional Hilbert space, as there is no constraint on the occupation number $n$ of a site. Each Hamiltonian in the direct-sum expansion now reads
\begin{align}
\label{E.Spectralb}
\hat{H}^{\left(n\right)}=\sum_{p_1,p_2,\dots,p_n=1}^N\left(\omega_{p_1}+\omega_{p_2}+\dots+\omega_{p_n}\right)\hat{c}_{p_n}^{\dagger}\dots\hat{c}_{p_2}^{\dagger}\hat{c}_{p_1}^{\dagger}\hat{c}_{p_1}\hat{c}_{p_2}\dots\hat{c}_{p_n}~,
\end{align}
with Hilbert space dimension $\text{dim}\left[\hat{H}^{\left(n\right)}\right]=\binom{N+n+1}{N-1}$, where $n$ are the number of bosons and $N$ the number of sites.

As, from Eqs.~\ref{E.Spectralf},\ref{E.Spectralb}, it is evident that the eigenenergies (eigenvectors) in the $n$-particle subspace are given simply by the sum (tensor product) of the single-particle eigenenergies (eigenvectors), in the next Section we will be able to write the many-body dynamics in terms of single-particle transition amplitudes.
\subsection{Many-body dynamics}\label{Ss.MBD}

Having sketched in the previous subsection the spectral decomposition of the Hamiltonian operator, it is easy now to express the dynamics of an arbitrary number of excitations in the chain in terms of single-particle dynamics~\cite{1367-2630-16-12-123003,1402-4896-2015-T165-014036,PhysRevLett.93.230502}.
The transition amplitude for the transfer of $n_s$ excitations, residing on the sender sites $\{n_s\}=\{s_1,s_2,\dots,s_{n_s}\}$, to the receiver sites $r$, residing on the receiver sites $\{n_r\}=\{r_1,r_2,\dots,r_{n_r}\}$, can be expressed in terms of the submatrix $\mathcal{F}_{\{n_s\}}^{\{n_r\}}(t)$ of the transition amplitude matrix $\mathcal{F}(t)$, where only the rows (columns) corresponding to the sites in the block $\{n_s\}$ ($\{n_r\}$) are taken into account. The transition matrix $\mathcal{F}(t)$ itself is built from single-particle transition amplitudes 
\begin{align}
\label{E.Sparticel_amp}
f_i^j(t)=\bra{j}e^{-it \hat{H}}\ket{i}=\sum_{k=1}^{N}e^{-i \omega_k t }\bra{j}\phi_k\rangle\langle \phi_k\ket{i}=\sum_{k=1}^{N}e^{-i \omega_k t }\phi_{jk}\phi^*_{ki}
\end{align}
as follows
\begin{eqnarray}\label{E.FMatrix}
\mathcal{F}(t)=
	\begin{pmatrix}
	f_1^1(t) & f_1^2(t) & \cdots & f_1^N(t)\\
	f_2^1(t) & \cdots & \cdots & f_2^N(t)\\
	\vdots & & \ddots &  \vdots\\
	f_N^1(t) &  & \cdots & f_N^N(t)\\
	\end{pmatrix}~,
\end{eqnarray}
where $\hat{H}$ is the Hamiltonian in Eq.~\ref{E_1part} and $\phi_k$ its eigenvectors.
Clearly $\mathcal{F}$ is unitary and, hence,
\begin{align}
\label{E.Conser}
\sum_{j=1}^{N}\left|f_i^j(t)\right|^2=1~,\forall i~ \text{and}~\sum_{i=1}^{N}\left|f_i^j(t)\right|^2=1~,\forall j~,
\end{align}
embody the normalisation condition for the single-particle transition probability from a fixed site index $i$, or to a fixed site index $j$, as expected by excitation number conservation.

As depicted in Fig.~\ref{F.Figure}, in the presence of the mirror symmetric Hamiltonian in Eq.~\ref{E_Ham1}, the eigenvectors
of the tridiagonal matrix $A$ are known to be either symmetric or antisymmetric~\cite{doi:10.1063/1.4797477}: $
\phi_{kn}=(-1)^{k+1}\phi_{k,N+1-n}$, with  $J_i>0$ and eigenvalues $\omega_k$ listed in decreasing order. 
This yields $f_i^j(t)=f_j^i(t)$ and $f_i^j(t)=f_{N+1-i}^{N+1-j}(t)$, resulting in a both persymmetric and centrosymmetric transition matrix $\mathcal{F}$. Clearly, once sender and receiver blocks (of the same size) are chosen at each edge of the chain, the resulting submatrix will retain only its persymmetry. Furthermore, the effect of a uniform potential $h$ on the eigenvalues $\omega_k$ in Eq.~\ref{E_1part} equals only to a uniform shift of their values at zero potential. As a result of the mirror-symmetry, the eigenvalues are symmetric around their middle value. Thus, one has $\omega_{\frac{N}{2}+i}=-\omega_{\frac{N}{2}+1-i}$, where $i=1,2,\dots,\frac{N}{2}$ for even $N$ and $\omega_{\frac{N+1}{2}+i}=-\omega_{\frac{N+1}{2}-i}$, where $i=0,1,2,\dots,\frac{N-1}{2}$ for odd $N$. All these conditions translate in having $f_i^j(t)$ purely real (imaginary) for even (odd) $i+j$.

In view of the previous results, we now explicitly construct the submatrix $\mathcal{F}_{\{n_s\}}^{\{n_r\}}(t)$ for an arbitrary number of excitations $n_s=n_r$.
Clearly, for $n_s\neq n_r$, the transition amplitude is identically null because of the excitation number conserving nature of the Hamiltonian.

Let us assume, without loss of generality, that each of the $n_s$ excitations resides on each site of the lattice at both edges, i.e, $\{n_s\}=1,2,\dots, n_s$ and $\{n_r\}=N+1-n_r,N+2-n_r,\dots,N$, see Fig.~\ref{F.Figure}.
Dropping henceforth the time-dependence and labeling by $n_s$ both the occupied set of sites and the number of excitations,
the relevant submatrix, $\mathcal{F}_{\{n_s\}}^{\{n_r\}}$ of  $\mathcal{F}$, is obtained by selecting the first $n_s$ rows and the last $n_r$ column,
\begin{eqnarray}\label{E.FsubMatrix1}
\mathcal{F}_{n_s}^{n_r}=
\begin{pmatrix}
f_1^{N+1-n_r}(t) & f_1^{N+2-n_r}(t) & \cdots & f_1^N(t)\\
f_2^{N+1-n_r}(t) & \cdots & \cdots & f_2^N(t)\\
\vdots & & \ddots &  \vdots\\
f_{n_s}^{N+1-n_r}(t) &  & \cdots & f_{n_s}^N(t)\\
\end{pmatrix}~.
\end{eqnarray}
Finally, the transition probability of $n_s$ excitations between the edges of the chain is obtained from the square modulus of the determinant  and the permanent of $\mathcal{F}_{n_s}^{n_r}$ for fermions and bosons, respectively. It is interesting to stress that, although in general $\left|\text{det}(\mathcal{F}_{n_s}^{n_r})\right|^2\neq \left|\text{perm}(\mathcal{F}_{n_s}^{n_r})\right|^2$, equality  is retrieved whenever all non-vanishing terms in the determinant have the same signature. This results, as we will show in the following, that at specific times -including when $\max_t\left[\mathcal{F}_{n_s}^{n_r}\right]$ is achieved- the transition probability of $n_s$ excitations between the edges of the chain is independent of its fermionic or bosonic nature.

In order to relax a bit the notation, hereafter we will label the $n_r$ receiver sites starting from the edge, $n_r=1,2,\dots, n_s$. This allows to highlight the persymmetry of the submatrix  $\mathcal{F}_{n_s}^{n_r}$
\begin{eqnarray}\label{E.FsubMatrix2}
\mathcal{F}_{n_s}^{n_r}=
\begin{pmatrix}
f_1^{n_s}(t) & f_1^{n_s-1}(t) & \cdots & f_1^1(t)\\
f_2^{n_s}(t) & \cdots & \cdots & f_2^1(t)\\
\vdots & & \ddots &  \vdots\\
f_{n_s}^{n_s}(t) &  & \cdots & f_{n_s}^1(t)\\
\end{pmatrix}~,
\end{eqnarray}
which now translates in $f_i^j(t)=f_j^i(t)$. As a consequence, there are only $\frac{n_s\left(n_s+1\right)}{2}$ distinct transition amplitudes in the submatrix in Eq.~\ref{E.FsubMatrix2}. Still, finding the conditions by which the transition probability approaches one is a formidable task. A determinant (permanent) of a $n_s$-dimensional square matrix is made up of a sum of $n_s!$ terms, each given by a product of $n_s$ transition amplitudes, of which, at most,  
$\ceil{\frac{n_s}{2}}$ terms are equal because of persymmetry, with  $\ceil{\bullet}$ being the ceiling function. Therefore, at least $\ceil{\frac{n_s}{2}}$ transition amplitudes have to reach one at the same time. Notice also that both $\text{det}\left(\mathcal{F}_{n_s}^{n_r}\right)$ and $\text{perm}\left(\mathcal{F}_{n_s}^{n_r}\right)$ are purely real (imaginary) for odd (even) lengths of the chain. Clearly $\mathcal{F}_{n_s}^{n_r}$ is not unitary, but $\left|\mathcal{F}_{n_s}^{n_r} \mathcal{F}_{n_s}^{{n_r}^{\dagger}}\right|_{ij}\leq 1$ and
\begin{align}
\label{E.Conser1}
\sum_{j\in\{n_r\}}\left|f_i^j(t)\right|^2\leq1~,\forall i\in \{n_s\}~ \text{and}~\sum_{i\in \{n_s\}}\left|f_i^j(t)\right|^2\leq1~,\forall j\in\{n_r\}~,
\end{align}
hold as a consequence of the particle-number conservation.

In this work, we derive the conditions for which the transition probability, both for fermions and bosons, approaches one by weakly coupling the sender and receiver block to the wire. We dub this dynamical regime PP-transfer (perturbatively-perfect transfer). In the following we set $J_0=0.01$, although we checked that PP-transfer does not depend on the specific value of $J_0$ insofar the weak-coupling condition $J_0\ll J=1$ is satisfied.

\subsection{PP-transfer}\label{Ss.PPtransfer}
Perturbative couplings have been used in several settings, from quantum-state transfer to entanglement generation. However, previous works focused mainly on one-excitation transfer~\cite{PhysRevA.72.034303, PhysRevA.78.022325, Zwick_2014}, with some exceptions dealing with two-excitation transfer~\cite{ doi:10.1142/S021974991750037X, 1402-4896-2015-T165-014036, PhysRevA.100.052308}. The case of $n>2$ excitation transfer has not yet been addressed in the perturbative regime. Let us first recap a few results for the one- and two-excitation PP-transfer which will be useful to describe the relevant dynamical features taking place also for $n_s>2$.

For one-excitation transfer, the bosonic or fermionic nature of the particle does not play any role, as there is no statistics involved.
\begin{figure}
	\label{F.Figure_1}
	\includegraphics[width =0.48\textwidth]{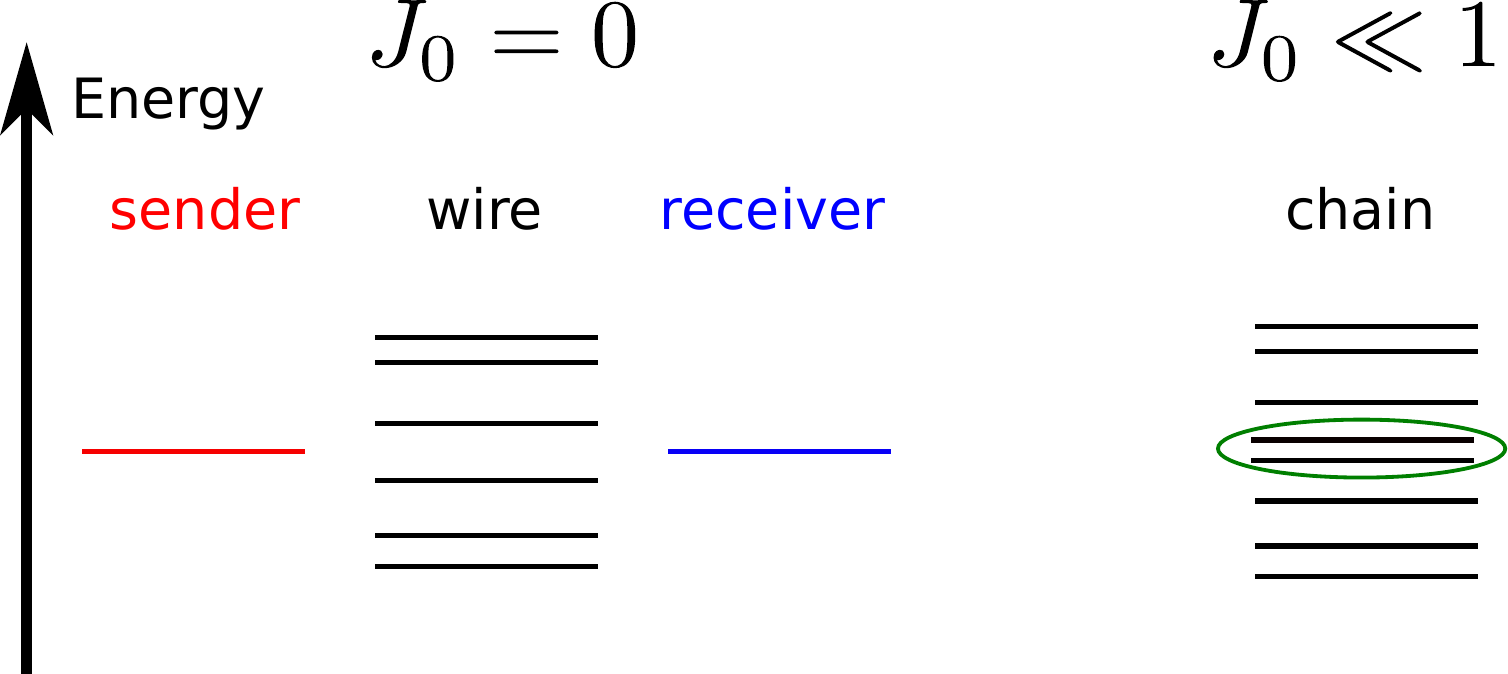}
	\vline
	\hspace{0.2cm}
	\includegraphics[width =0.48\textwidth]{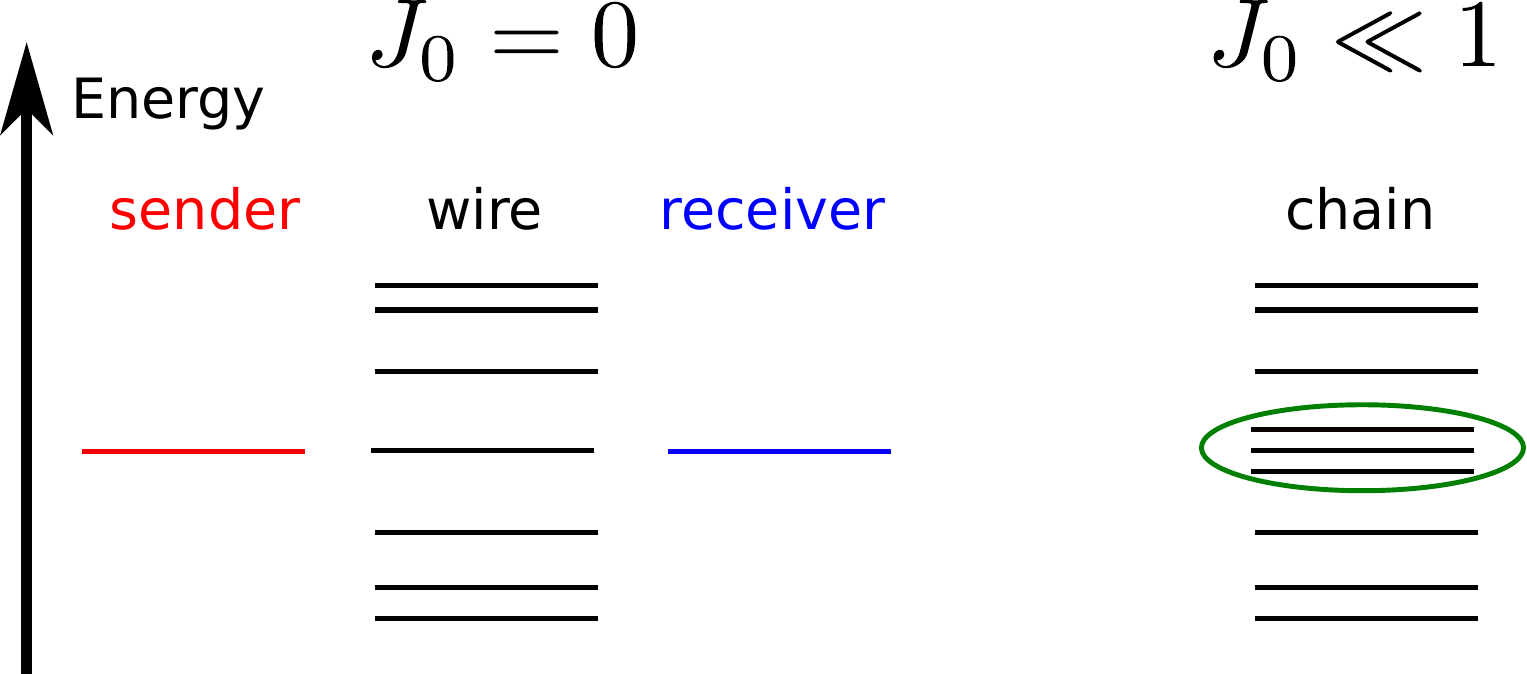}
	\caption{Single-particle energy spectrum of the chain, composed by one sender, one receiver, and a wire of even and odd length $n_w$, respectively, left and right panel. In both cases, the energy levels, before and after the coupling is switched on, are shown. Notice that, for an even length of the wire (left panel) no resonances occur between the sender (receiver) eigenenergy and the wire's one, at variance with the case of an odd length of the wire (right panel). As a consequence, for $n_w$ even, two quasi-degenerate eigenenergies, whose eigenstates are localised on the sender and receiver site, enter in Eq.~\ref{E.Sparticel_amp}. For $n_w$ odd, instead, a wire energy eigenstate is resonant with the sender (receiver) energy and three quasi-degenerate eigenstates enter the dynamics depicted by Eq.~\ref{E.Sparticel_amp}. This yields, in the latter case, a non-zero probability for the excitation to be found in the wire. The green line highlights the quasi-degenerate states. }
\end{figure}
For $n_s=1$, the transfer amplitude, 
is given simply by Eq.~\ref{E.Sparticel_amp}. Because of the perturbative coupling, only the two (three) eigenvectors, lying in the middle of the single-particle spectrum, have non-negligible overlap with the initial and final state, see Fig.~\ref{F.Figure_1}. This reduces the transition probability to

\begin{align}
\label{E.1Ex}
\left|f_1^N(t)\right|^2=
\begin{cases}
&\left|\sum_{k=\frac{N}{2}}^{\frac{N}{2}+1}e^{-i \omega_k t }\phi_{Nk}\phi^*_{k1}\right|^2 = \left|\left|\phi_{\frac{N}{2}1}\right|^2\left(1-\cos \omega_e t\right)\right|^2 \qquad \text{for $N$ odd}\\
\\
&\left|\sum_{k=\frac{N+1}{2}-1}^{\frac{N+1}{2}+1}e^{-i \omega_k t }\phi_{Nk}\phi^*_{k1}\right|^2=\left|2 i \left|\phi_{\frac{N+1}{2}-1, 1}\right|^2 \sin \omega_o t\right| \qquad \text{for $N$ even}
\end{cases}~,
\end{align} 
where, for even $N$, $\omega_e=\frac{E_{\frac{N}{2}+1}-E_{\frac{N}{2}}}{2}\sim J_0^2$ and $\phi_{\frac{N}{2},1}\simeq \frac{1}{\sqrt{2}}$, and, for odd $N$,  $\omega_e=\frac{E_{\frac{N+1}{2}}-E_{\frac{N+1}{2}-1}}{2}\sim J_0$, and $\left|\phi_{\frac{N+1}{2}-1, 1}\right| \simeq \frac{1}{\sqrt{2}}$ only for $N\gg 1$. The approximate values for these coefficients can be obtained by a simple procedure, which we illustrate below for even $N$. When $J_0=0$, the sender and the receiver each have one eigenenergy state in the single-excitation sector with energy $E=h$, that $\ket{1}$ and $\ket{N}$, respectively. In the presence of a perturbative coupling, $J_0\ll 1$, the degeneracy between the sender and the receiver eigenstate is broken and, in the single-particle sector, the eigenstates are $\ket{\Psi^\pm}\simeq \frac{1}{\sqrt{2}}\left(\ket{0N}\pm \ket{10}\right)$, because of mirror-symmetry. From Eqs.~\ref{E.1Ex}, we see that excitation transfer can be achieved with a probability perturbatively close to one.
A similar procedure yields PP-transfer of one excitation also for odd $N$.
The main difference between the two cases lies in the fact that, for even $N$, there are no resonances between the sender (receiver) and the wire single-particle energy states, whereas, for odd $N$ such a resonance occurs, see Fig.~\ref{F.Figure_1}. Consequently, in the former case the energy splitting is a second-order perturbative effect, whereas, in the latter, it is a first-order one. This translates in shorter transfer times for the odd-length wire.

The characteristic feature of 1-excitation transfer in Eqs.~\ref{E.1Ex} is the presence of a single frequency, which gives rise to Rabi-like oscillations of the excitation between the pair of two-level systems embodied by the sender and the receiver qubit. For $n=1$ excitation transfer, this is a direct consequence of the weak-coupling which couples perturbatively only two (three) single-particle levels. On the other hand, for $n>1$, there will be  more levels entering the dynamics (the precise number will be given in the next subsection) and, therefore,  more frequencies enter the sum of the transition amplitude in Eq.~\ref{E.Sparticel_amp}. As a consequence, Rabi-like oscillations are much harder to achieve. Nevertheless, if one of the frequencies is much smaller than every other, then it will dominate the dynamics, i.e., it will form the envelop of the transition amplitude in Eq.~\ref{E.Sparticel_amp} and, therefore, unit probability is achievable in a Rabi-like dynamical scenario. Such a scenario is here defined as PP-transfer. 
Let us also specify that here we are referring to PP-transfer of excitations, that is, having the determinant (permanent) of Eq.~\ref{E.FsubMatrix2} equal to one. Although every physical quantity in quadratic models can be expressed in terms of single-particle amplitude, PP-transfer of excitations does not necessarily imply PP-transfer of, say, an arbitrary quantum state. Nevertheless, as we will see in Sec.~\ref{S.Applications}, PP-transfer of excitations implies PP-transfer of energy and magnetisation, for instance.


\subsection{Resonances in sender-wire-receiver system}\label{Ss.res}
In order to determine the number of eigenstates giving a non-negligible contribution to the transition amplitude in Eq.~\ref{E.Sparticel_amp}, it is necessary to identify which states of the sender (receiver) block exhibit resonances with the wire's eigenstates.
In the weak-coupling regime, this identifies  different lengths of the wire giving rise to resonances between its eigenenergies and those of the sender (receiver) block. 

As shown in the previous subsection, for an excitation sitting initially on the first site, $\ket{1}$, only two or three terms are relevant in the wave packet in Eq.~\ref{E.Sparticel_amp}, depending whether $N$ is even or odd, respectively~\cite{PhysRevA.72.034303}. In the former case there are only two eigenvectors $\ket{\phi_k}$ of the system having a non-negligible overlap with sites $1$ and $N$, whilst in the latter they amount to three. This can be easily deduced by considering the number of resonant energy levels of the uncoupled system, sender, receiver, and wire. For $J_0=0$, there is only one single-particle energy eigenstate for the sender and the receiver, respectively, with energy $E=h$. The energy spectrum of the wire is given by $E_k=h+\cos\frac{k \pi}{n_w+1}$~\cite{PhysRevA.79.022302}. Therefore, in order to have degeneracy between the sender (receiver) and the wire, $N$ has to be odd as the condition $\frac{k \pi}{n_w+1}=0$ has to hold. When $J_0$ is switched on in the weak-coupling limit, $J_0\ll 1$, the degeneracy is lifted by $\delta$. For even $N$, it becomes a second-order perturbation effect, and the energy splitting is $\Or(J_0^{2})$, whereas, for odd $N$, the effect is of first order yielding an energy splitting $\Or(J_0)$. Being the transfer time $\tau\propto \delta^{-1}$, PP-transfer in odd-length chains is  faster than in even-length ones. 

Now we consider the case of $n_s=n_r>1$. In order to have resonant energy levels with the wire, made of $n_w$ sites, the following condition has to hold
\begin{align}
\label{E.resonant_condition}
\frac{k \pi}{n_s+1}=\frac{q \pi}{n_w+1}~,~k=1,..,n_s~\text{and}~q=1,..,n_w~.
\end{align}
which, when put in the following form
\begin{align}
\label{E.resonant_condition2}
q=\frac{n_w+1}{n_s+1}k~,
\end{align}
shows that whenever two length of wire, $n_w$ and $m_{w}$ are congruent modulo $n_s+1$, i.e.,  $n_w\equiv m_{w}~\left(\text{mod}\left(n_s+1\right)\right)$, 
the two wires share the same number of resonant modes with the sender. As a consequence, different lengths of wire $n_w$, but belonging to the same equivalence class, will exhibit similar dynamical behaviour, in particular, with respect to PP excitation transfer.
To find the number of resonant modes $n_{res}$, one simply solves Eq.~\ref{E.resonant_condition2} for each integer $p$ in the least residue system $\text{mod}\left(n_s+1\right)$, i.e., $p=0,1,\dots,n_s$. It turns out that the mode $q$ of the wire is resonant with the mode $k$ of the sender for
\begin{align}
\label{E.resonant_condition3}
q=\frac{m\left(n_s+1\right)+p+1}{n_s+1}k=\left(m+\frac{p+1}{n_s+1}\right)k~,
\end{align}
where $m$ is an integer and the length of the wire is $n_w=m\left(n_s+1\right)+p$.

A few instances, relevant in the following, will be  analysed. For  $p=0$, hence a wire of length $n_w=m\left(n_s+1\right)$, Eq.~\ref{E.resonant_condition3} reads
\begin{align}
\label{E.resonant_condition4}
q=mk+\frac{1}{n_s+1}k~.
\end{align}
This equation never holds as $k<n_s+1$ and therefore no resonances are present between the wire and the sender for arbitrary $n_s$. For $p=1$ and  $n_w=m\left(n_s+1\right)+1$, one gets
\begin{align}
\label{E.resonant_condition5}
q=mk+\frac{2}{n_s+1}k~,
\end{align}
which is satisfied only for  $n_s$ odd and brings about resonance between the mode $k=\frac{n_s+1}{2}$ and $q=\frac{n_w-1}{2}+1$ of the sender block and the wire, respectively. Furthermore, because of the reflection symmetry of the energy spectrum of both systems, this is the only resonance present. Finally, we consider the case $p=n_s$, corresponding to $n_w=m\left(n_s+1\right)+n_s$. Eq.~\ref{E.resonant_condition3} becomes $q=\left(m+1\right)k$, meaning that each sender energy eigenstate is resonant with one eigenstate of the wire. This is the maximum number of resonances in the system as the energy levels of the uncoupled blocks in Fig.~\ref{F.Figure} are non-degenerate. 

Following such a procedure for each $p$ it is easy to build the  table~\ref{table:nonlin} for an arbitrary number of senders $n_s$.
\begin{table}[h!]
	\centering
\begin{tabular}{|*{15}{c|}}
		\hline
		$n_s$ & \multicolumn{2}{c|}{1} &  \multicolumn{3}{c|}{2} & \multicolumn{4}{c|}{3} & \multicolumn{5}{c|}{4} \\
		\hline
		 $n_w$ &\multicolumn{2}{c|}{$2l$ $2l+1$}&\multicolumn{3}{c|}{$3l$ $3l+1$ $3l+2$}& \multicolumn{4}{c|}{$4l$ $4l+1$ $4l+2$ $4l+3$} &\multicolumn{5}{c|}{$5l$ $5l+1$ $5l+2$ $5l+3$ $5l+4$}\\
		\hline
		$n_{res}$ &\multicolumn{2}{c|}{0 1}&\multicolumn{3}{c|}{0 0 2}& \multicolumn{4}{c|}{0 1 0 3} &\multicolumn{5}{c|}{0 0 0 4}\\
		\hline
	\end{tabular}
\label{table:nonlin}
	\caption{Table showing the number of resonances between the sender and the wire $n_{res}$ for $n_s$ senders up to 4 and a wire of length $n_w$, with $l=0,1,2,\dots$.}
\end{table}

\section{Many-body dynamics}\label{Ss.MBD1}
Now, before dealing with the case $n_s>2$, we first discuss some of the results obtained in Ref.~\cite{doi:10.1142/S021974991750037X} for the case of two-excitation transfer. Our previous discussion about the perturbation order of the sender-wire resonances immediately explains the reason wires of length $n_w=3l+2$ perform PP quantum-state transfer in a faster time than wires of length $n_w\neq 3l+2$. Indeed, the former case exhibits first-order perturbation correction to the three-fold quasi-degenerate energy eigenstates relevant to Eq.~\ref{E.Sparticel_amp}, whereas, in the latter case, the first correction to the two-fold quasi-degenerate eigenstates is of second-order.
For the details about the transfer time and the perturbative expansions we refer the reader to Ref.~\cite{doi:10.1142/S021974991750037X}, and for the generation of entangled states between the sender and receiver block to Ref.~\cite{PhysRevA.100.052308}. 

\begin{figure}
	\label{F.2ex_bos_fer}
	\includegraphics[width=\textwidth]{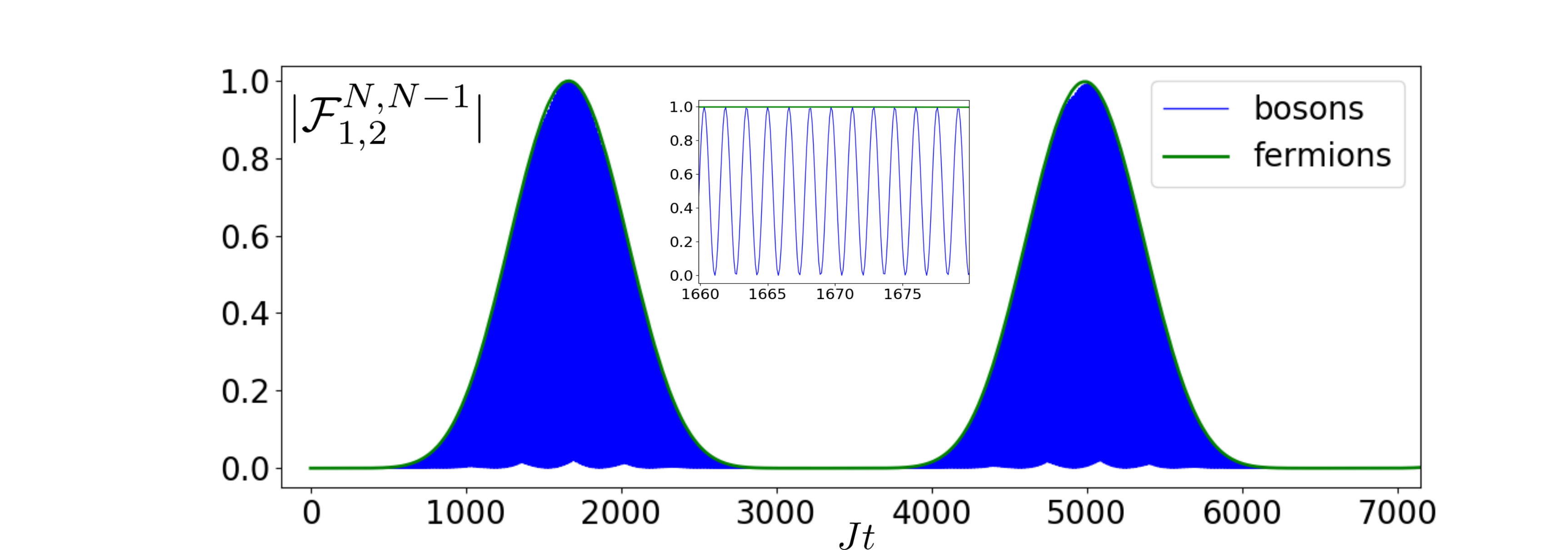}
	\caption{The transition probability of two excitations from sender sites $\{n_s\}=1,2$ to receiver sites $\{n_r\}=N-1,N$ for fermions and bosons, respectively. Excitations bounce back and forth between the sender and the receiver block via a Rabi-like dynamics, where the green curve is for fermions and constitutes the envelop of the blue curve for bosons. The inset shows the dynamics in a interval of around unit transfer probability for fermions, highlighting the boson dynamics on a time scale of order of $J$. The length of the chain is $N=45$.}
\end{figure}

Here we highlight the fact that the bosonic or fermionic nature of the excitations plays a key role in the dynamics because of the different dimensions of the Hilbert space of the Hamiltonian in Eq.~\ref{E_Ham1} due to their different statistics. Indeed, as for fermions the receiver's Fock space is made up of a single state in the two-particle sector, namely $\ket{11}$, for bosons, in addition to the latter, also the states $\ket{02}$ and $\ket{20}$ build up the Fock space. Consequently, the transition probability between the states $\ket{12}$ and $\ket{N{-}1~N}$ for fermions and bosons are not equivalent at all times. Nevertheless, the fermionic transition amplitude envelops the bosonic one, with the two bosons exploring the receiver's Hilbert space on a time scale $J$, see Fig.~\ref{F.2ex_bos_fer}. It is worthwhile to anticipate that such a difference of their respective transition probabilities does not have consequences on several observables, such as the average excitation number on each site, as we will show in Sec.~\ref{S.Applications}.

\subsection{Equivalence between bosonic and fermionic PP-excitations transfer}
As can be seen from Fig~\ref{F.2ex_bos_fer}, the two-fermion transition probability is the envelop for the bosonic one. As a consequence, PP-tranfer is achieved at the same time for both classes of particles. This is a general feature of the model and  can be explained by means of perturbation theory.

In the weak-coupling limit, and in the absence of resonances with the wire, we can approximate the perturbed eigenstates having non-zero overlap with the sender and the receiver sites as the symmetric and antisymmetric linear combination of the degenerate single-particle eigenstates of the sender and the receiver block
\begin{align}
\label{E.symmetric_comb_1}
\ket{\Psi_k}_{sr}=\frac{1}{\sqrt{2}}\left[\sqrt{\frac{2}{n_s+1}}\sum_{l=1}^{n_s}\sin\frac{k \pi l}{n_s+1}\ket{l}\pm \sqrt{\frac{2}{n_s+1}}\sum_{l=1}^{n_s}\sin\frac{k \pi l}{n_s+1}\ket{N+1-l} \right]~.
\end{align}
It turns out that the transition amplitude in Eq.~\ref{E.Sparticel_amp} is bounded by
\begin{align}
\label{E_bound}
\max_{t}\left|f_i^j(t)\right|=\max_{t}\left|\sum_{k=1}^{N}e^{-i \omega_k t }\phi_{jk}\phi^*_{ki}\right|\leq\ \sum_{k=1}^{N}\left|\phi_{jk}\phi^*_{ki}\right|=\frac{2}{n_s+1}\sum_{k\in\{n_s\}}\left|\sin \frac{k \pi j}{n_s+1}\sin \frac{k \pi i}{n_s+1}\right|~.
\end{align}
The last term on the RHS of Eq.~\ref{E_bound} is equal to one only for $i+j=n_s+1$ and $i=j$. This translates in a submatrix $\mathcal{F}_{n_s}^{n_r}$ (Eq.~\ref{E.FsubMatrix2}) which can, at most, have unit single-particle amplitudes either on the main diagonal or on the skew-diagonal, respectively. Although, the determinant (permanent) of $\mathcal{F}_{n_s}^{n_r}$ may become one also due the contribution of many terms in their respective expansion, it is highly improbable that such a coherent fully constructive interference between wavepackets $f_i^j(t)$, for arbitrary $i$ and $j$, will take place, especially in the presence of many frequencies entering the dynamics. As a result, also in view of the normalisation condition in Eq.~\ref{E.Conser}, PP-transfer is most likely to occur when all the terms either on the main, or on the skew-diagonal, will reach unit single-particle transition amplitude. It is now immediate to realise that PP-transfer for an arbitrary number of excitations is independent from their bosonic or fermionic nature as the signature of the determinant of Eq.~\ref{E.FsubMatrix2} does not play any role.

\subsection{Heuristic approach to PP transfer}
The result in Eq.~\ref{E_bound} and the resonance conditions derived in Sec.~\ref{Ss.res} allow us also to give a rule of thumb as to whether PP-excitation transfer is achievable for an arbitrary number of excitations $n_s$ in a wire of given length $n_w$ by means of the protocol of Fig.~\ref{F.Figure}. Building on the argument for one-particle transfer in Sec.~\ref{Ss.PPtransfer}, the single-particle transition amplitudes entering the submatrix in Eq.~\ref{E.FsubMatrix1} are given by Eq.~\ref{E.Sparticel_amp}, where only resonant modes have to be kept. 

Let us consider the case where $j=N+1-i$, i.e., mirror-symmetric sites in the sender and receiver block, respectively, corresponding to the elements on the diagonal of the matrix in Eq.~\ref{E.FsubMatrix2}. 
In the absence of resonant modes with the wire, Eq.~\ref{E.Sparticel_amp} reads
\begin{align}
\label{E.Sparticel_amp_gen_1}
f_i^j(t)&=\sum_{k=1}^{2n_s}e^{-i \omega_k t }\phi_{jk}\phi^*_{ki}=\sum_{k=1}^{2n_s}\phi_{jk}\phi^*_{ki}\left(e^{-i \delta_k t }- e^{i \delta_k t }\right)e^{-i \omega_k t }\nonumber\\
&=\sum_{k=1}^{n_s}\phi_{jk}\phi^*_{ki}\left(e^{-i \delta_k t }- e^{i \delta_k t }\right)\left(e^{-i \omega_k t }+ e^{i \omega_k t }\right)=\sum_{k=1}^{n_s}\frac{\phi_{ik}\phi^*_{ki}}{4}\sin \delta_k t\cos\omega_k t~,
\end{align}
where in the last line, without loss of generality, we have considered an instance of even $i+j$ and mirror-symmetry of the energy spectrum has been exploited. The transition amplitude is hence given by a wavepacket of $n_s$ travelling waves, each given by a product of harmonic functions, being sines or cosines depending on $n_w$, $n_s$, and $k$. The specific form of the harmonic function of $\omega_k$ not being relevant, we notice that the frequencies entering the functions satisfy $\delta_k \ll \omega_k$, as the energy shift of the $k$-th energy level is negligible with respect to its unperturbed value. As a consequence, $\sin \delta_k t$ shapes up the envelop of $k$-th wave of $f_i^j(t)$. Therefore it is straightforward to conclude that, in order to have $f_i^j(t)=1$ at some specific time $t=\tau$, the $\delta_k$'s should be all commensurate, which is a hard condition to fulfill, or only one $\delta_k^*$ should be much smaller than all the others. The latter condition defines the rule of thumb for PP-excitation transfer:
\begin{align}
\label{E_ruleofthumb}
\exists!~\delta_k \ll \delta_q~,
\end{align}
where the $\delta$'s are the energy shifts of the corresponding energy levels entering Eq.~\ref{E.Sparticel_amp_gen_1}. Eq.~\ref{E_ruleofthumb} states that if in the wavepacket of Eq.~\ref{E.Sparticel_amp_gen_1} there is only one energy being corrected at an higher order in perturbation theory, PP excitation transfer is attainable.
Indeed, being $n$-excitation transfer achievable by the product of the single-particle transfer on the (skew) diagonal, each evolving with the same eigenenergies as in Eq.~\ref{E.Sparticel_amp_gen_1}, the transfer time is simply given by $\tau\simeq\frac{\pi}{2 \delta_k^*}$. 

An identical argument applies in the presence of resonances with the wire where the single-particle transition amplitude reads
\begin{align}
\label{E.Sparticel_amp_gen_2}
f_i^j(t)=\sum_{k=1}^{2n_s+n_w}e^{-i \omega_k t }\phi_{jk}\phi^*_{ki}
\end{align}
and the energy shifts $\delta_k$ are evaluated taking into account the triple quasi-degenerate nature of the energy level(s).

\subsection{3- and 4-excitation PP transfer}
Let us now address the case of $n_s>2$. 
For three  fermionic excitations, in order to have PP transfer
\begin{align}
\label{E.3ampl}
|\mathcal{F}_s^r|^{2} = 
\begin{vmatrix}
f_{1}^{N-2}& f_{1}^{N-1} &  f_{1}^{N}\\
f_{2}^{N-2} &f_{2}^{N-1} & f_{1}^{N-1}\\
f_{3}^{N-2} &f_{2}^{N-2} &  f_{1}^{N-2}\\
\end{vmatrix}^2 \simeq 1~,
\end{align}
where, without ambiguity, we have labelled by $s$ and $r$ the sender and receiver sites, respectively. 
According to the arguments in the previous sections, we analyse the contribution of main diagonal to the determinant, with similar arguments holding for the skew diagonal contribution, 
\begin{align}
\label{E.3ampl.2}
|\mathcal{F}_{s}^{r}|^{2} = \left|\left(f_1^{N-2}\right)^2 f_2^{N-1}\right|^2~.
\end{align}

For the case of $n_w=4n+1$, the single-particle transition amplitude in Eq.~\ref{E.Sparticel_amp} now reads
\begin{align}
\label{E.3_1_Res}
f_i^j(t)=\sum_{k=1}^7 e^{-i \omega_k t }\phi_{jk}\phi^*_{ki}~.
\end{align}
From Table~\ref{table:nonlin} we notice that two double quasi-degenerate  and one triple quasi-degenerate eigenstates have non-negligible overlap with the sender and receiver sites. As the former degeneracy is resolved at second-order in perturbation theory, and the latter at first-order, this implies that, for $J_0\rightarrow 0$, we may expect the rule of thumb in Eq.~\ref{E_ruleofthumb} to hold as \nth{2}-order energy shifts are $\Or(J_0^{2})$ wheareas \nth{1}-order shifts are $\Or(J_0)$.

Indeed, we see that PP transfer is ruled by the following term
\begin{align}
\label{E.3vaenv}
|\mathcal{F}_{s}^{r}(t)|\simeq\left| \sin^2\omega_{76}^-t \right|^2~,
\end{align}
where $\omega_{76}^-=\frac{E_7-E_6}{2}$ is the \nth{2}-order perturbation energy shift of the double quasi-degenerate energy eigenstate. The positions of $E_6$ and $E_7$ of Eq.~\ref{E_1part} in the single-particle energy spectrum of the chain, ordered in increasing values, are given by $k=\floor{ \frac{N+1}{2}\cos^{-1}\frac{1}{\sqrt{2}}}-1$ and $\floor{ \frac{N+1}{2}\cos^{-1}\frac{1}{\sqrt{2}}}-2$, respectively.

Concerning the other lengths of wire $n_w$ in Table~\ref{table:nonlin}, for $n_s=3$, we notice that they all have exclusively \nth{1}- or \nth{2}-order perturbation energy corrections. By the rule of thumb in Eq.~\ref{E_ruleofthumb}, we do not expect PP transfer, being all the energy shifts of the same order of magnitude for a given $n_w$. In addition, we  show that also non-PP transfer does not occur, being the energy shifts incommensurate.

Let us first analyse the non-resonant cases in Table~\ref{table:nonlin} $n_w=4n,4n+2$. As only six eigenstates take part in the dynamics, the single particle transition amplitude between a sender and a receiver site reduces to
\begin{align}
\label{E.3nonRes}
f_i^j(t)=\sum_{k=1}^6 e^{-i \omega_k t }\phi_{jk}\phi^*_{ki}~.
\end{align}
From the perturbative expansion of Eq.~\ref{E.symmetric_comb_1}, the envelop of the transition amplitude in Eq.~\ref{E.3nonRes} can be written as
\begin{align}
\label{E.3nonva}
|\mathcal{F}_{s}^{r}(t)|=\left|\frac{1}{4}\left(\sin E_4t+\sin\frac{E_6-E_5}{2}t\right)^2 \sin\frac{E_6-E_5}{2}t\right|~,
\end{align}
where $E_4$ is given by the energy level labeled by $k=\frac{N}{2}+1$, $E_6$ and $E_5$ by $k=\floor{ \frac{N+1}{2}\cos^{-1}\frac{1}{\sqrt{2}}}$ and $\floor{ \frac{N+1}{2}\cos^{-1}\frac{1}{\sqrt{2}}}+1$, respectively.
From Eq.~\ref{E.3nonva}, it is evident that, in order to achieve transfer of 3 excitations, $E_4$ and $E_6-E_5$ have to be commensurate. This implies that 
\begin{align}
\label{E.3comm}
\begin{cases}
E_4 t&=\frac{\left(4n+1\right)\pi}{2}\\
\frac{E_6-E_5}{2}t&=\frac{\left(4m+1\right)\pi}{2}
\end{cases}~\text{and}~
\begin{cases}
E_4 t&=\frac{\left(4n+3\right)\pi}{2}\\
\frac{E_6-E_5}{2}t&=\frac{\left(4m+3\right)\pi}{2}
\end{cases}~,
\end{align}
have to hold with $n$ and $m$ integers,
in order to have the oscillatory functions in Eq.~\ref{E.3nonva} be 1 or -1 at the same time.
Hence, one of the two following conditions has to be fulfilled
\begin{align}
\label{E.cond}
\frac{E_6-E_5}{2E_4}=\frac{4m+1}{4n+1}~\text{and}~\frac{E_6-E_5}{2E_4}=\frac{4m+3}{4n+3}~.
\end{align}
The impossibility of the transfer arises because, for $J_0\rightarrow 0$, we find numerically that the energy ratio $\frac{E_6-E_5}{2E_4}\rightarrow \frac{1}{2}$.
Therefore, Eqs.~\ref{E.cond} can not be fulfilled by any integer pair $n$ and $m$, as can be readily seen from the fact that they can be cast into 
\begin{align}
\label{E.cond1}
8m=4n-1~\text{and}~8m=4n-3~,
\end{align}
respectively. 
The same argument about incommensurability of the eigenfrequencies entering Eq.~\ref{E.3nonRes} applies for wires of length  $n_w=4n+3$. Notice that, in the latter case, according to Table~\ref{table:nonlin} there are 3 sets of triple quasi-degenerate eigenstates, all coming from \nth{1}-order perturbation expansion. Nevertheless the same argument applies as the ratio of the energy shifts is found numerically to be $\frac{1}{2}$ for $J_0\rightarrow 0$. 

Clearly, as we are reporting a limiting procedure, there may be instances of $J_0$ where the ratio becomes quasi-commensurate, and after a very large amount of time a transfer probability close to one may be achieved. Such fortuitous cases, however, are not the topic of our investigation, as we are considering the conditions to be fulfilled in order to achieve PP-transfer in the generic limit of weak coupling instead of some specific values of $J_0$, which may eventually be a set of zero measure and hence extremely sensible to disorder. 

\begin{figure}
	\centering
	\includegraphics[width=\linewidth]{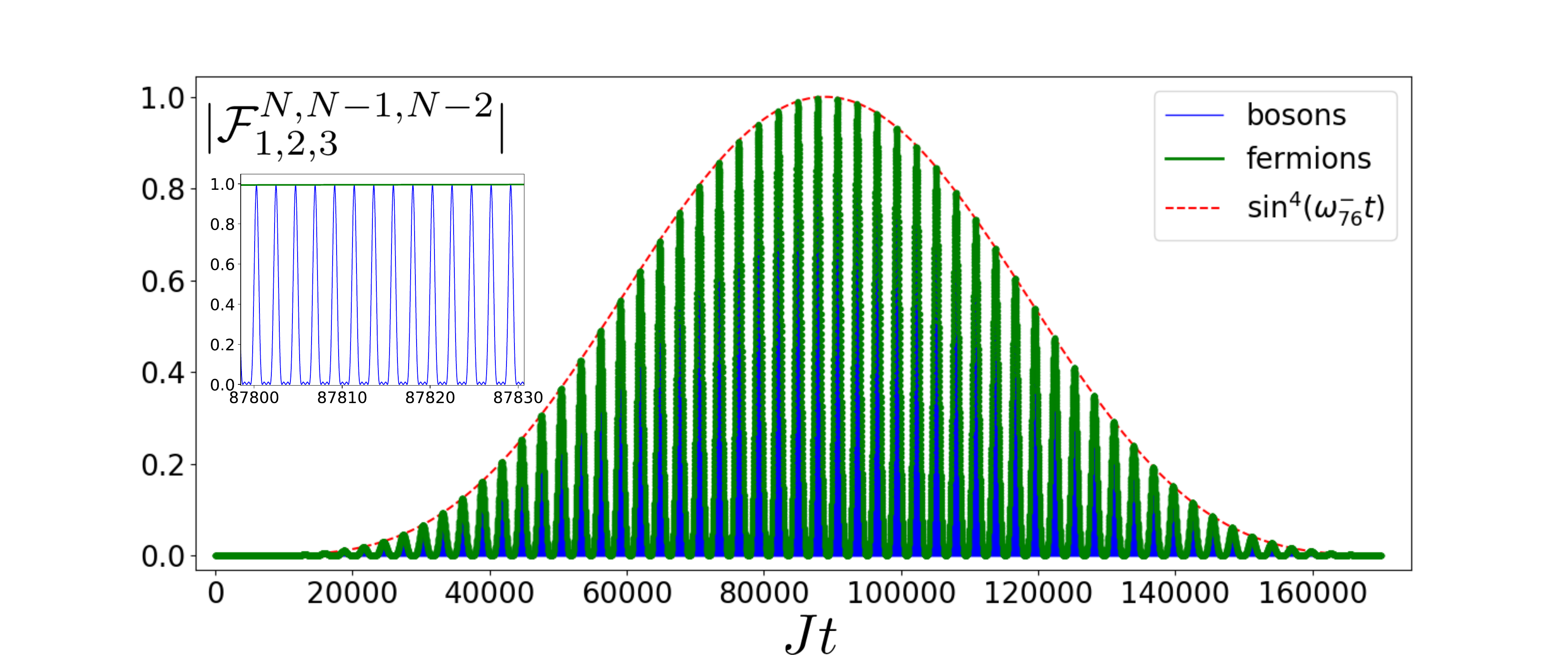}
	\caption{Transition probability for the transfer of three excitations from sites 1,2,3 to sites $N-2,N-1,N$ for a chain length of $N=47$ in a time interval of $t\in [0,160000/J]$. The blue (green) line shows the exact dynamics for bosons (fermions), whilst the red dotted line is the envelop of the three-particle transition as calculated in Equation~\eqref{E.3vaenv}. In the inset, a zoom around the time of PP transfer is shown.}
	\label{fig:dyn3}
\end{figure}

To summarise, we have found that for $n_s=3$ excitations, placed at one edge of a wire of length $n_w$ and in the weak-coupling limit $J_0\rightarrow 0$, PP-transfer is achievable only for $n_w=4l+1$ where the unique \nth{2}-order perturbation eigenenergy correction determines the transfer time. Other equivalence classes of the wire's length do not achieve unit transfer of three excitations because all the energy shifts belong to the same perturbation order and commensurability between frequencies is not achieved. In Fig.~\ref{fig:dyn3} we depict the results only for the case of successful PP-transfer.  

Let us now address the case of $n_s=4$. From Table~\ref{table:nonlin} we see that all energy shifts, for a given $n_w$, are of the same order in perturbation theory, either \nth{1}-order for $n_w=5l+4$ or \nth{2}-order in all the other case. Therefore, at variance with the case $n_s=3$, the condition for PP-excitation transfer given by Eq.~\ref{E_ruleofthumb} is not satisfied. Nevertheless, there are lengths of the wire $n_w$ exhibiting successful $n_s=4$ excitation transfer, whereas other lengths do not. The reason, as we will show, can be traced back to the fact that for some length of wires $n_w$, the energy splitting at \nth{2}-order in perturbation theory is almost one order of magnitude lower for some energy levels than it is for others. Let us analyse first the successful case. 

\begin{figure}
	\centering
	\includegraphics[width=\linewidth]{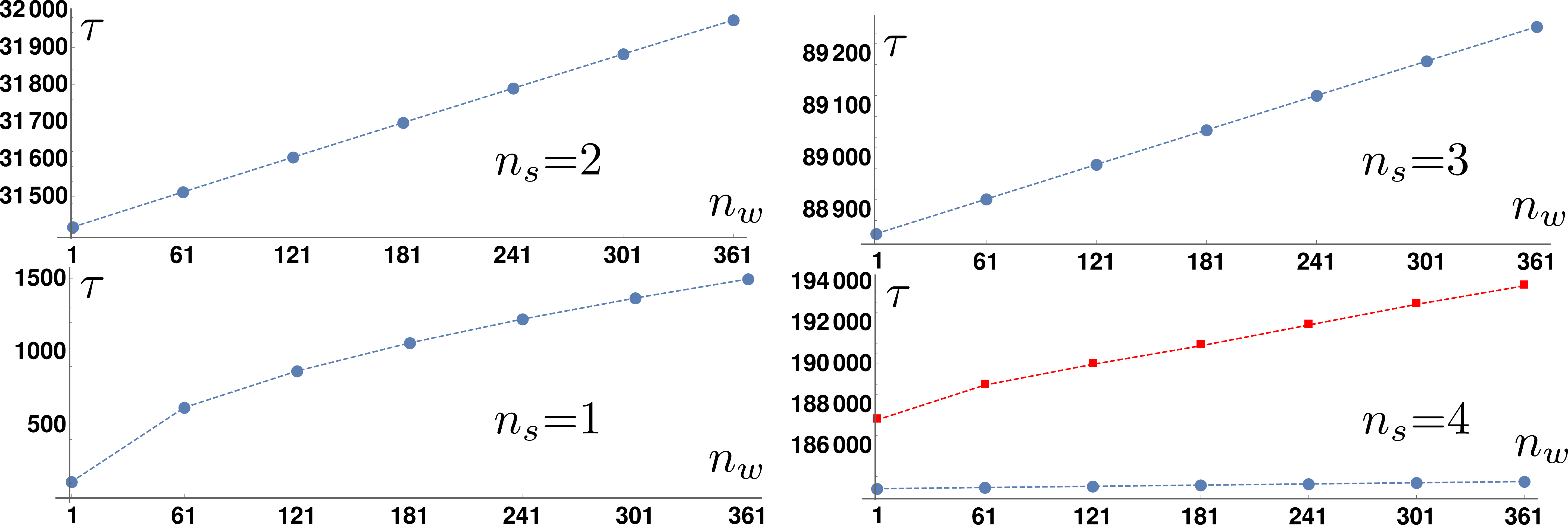}
	\caption{Transfer times $\tau$  for $n_s=1,2,3,4$ excitations in wires of different length $n_w$ fulfilling the (quasi) PP condition for all considered $n_s$. Notice that, whilst for $n_s=1$ the transfer time increases as $\sqrt{n_w}$ because the relevant frequency is obtained by first-order perturbation theory, in all other cases the slowest frequency is given by second-order perturbation theory, yielding thus a linear increase with $n_w$. In the lower right panel, for $n_s=4$, the blue curve is the exact numeric transfer time, whereas the red one reports $\tau=\frac{\pi}{\omega_{78}^-}$. Both follow a linear increase, although with different slope.}
	\label{fig:tim3rm}
\end{figure}

For $n_w=5l+2$, the perturbated eigenstates are located at position $k=\floor{ \frac{N+1}{\pi}\cos^{-1}\frac{\sqrt{5}+1}{4}}-1$ and $k=\floor{ \frac{N+1}{\pi}\cos^{-1}\frac{\sqrt{5}+1}{4}}$ for the higher energy state, and $k=\floor{ \frac{N+1}{\pi}\cos^{-1}\left(\frac{\sqrt{5}-1}{4}\right)}$ and $k=\floor{\frac{N+1}{\pi}\cos^{-1}\left(\frac{\sqrt{5}-1}{4}\right)}+1$ for the lower one. By numerical evaluation, we obtain that the ratio $\omega_{78}^-$ to $\omega_{56}^-$ goes to 0.14, for $J_0\ll 1$ and irrespective of $l$. The same situation occurs for $n_w=5l+1$. In these cases $n_s=4$ excitation transfer occurs, although it is not ruled by a single frequency and hence, according to our definition, is is not PP-excitation transfer. Indeed, in order to determine the transfer time, one has to find the maximum of two-single particle transition amplitudes entering the 4 excitation transition probability between the edges of the chain,
\begin{align}
\label{E.4ex}
\left|\mathcal{F}_s^r\right|^{2}=\left|\left(f_1^{N-3}\right)^2\left(f_2^{N-2}\right)^2\right|^2~.
\end{align}
Clearly, the fact that one energy shift is almost one order of magnitude lower than the other entering the dynamics, allows to determine the order of magnitude of the transfer time as given by $\tau=\frac{\pi}{2 \omega_{78}^-}$. In Fig.~\ref{fig:tim3rm} a comparison of the latter with the exact numerical result for excitation transfer is shown in panel d.
On the other hand, for $n_w=5l, 5l+3, 5l+4$, one has $\omega_{56}^-\simeq \omega_{78}^-$, with the ratio going to 0.38. PP-transfer does not occur and also non-PP transfer has not been found for several instances within time intervals related to the inverse of the energy splits. Clearly, this does not mean that the excitations may not be transferred at a certain time, being only two frequencies involved and occasional instances of commensurability may occur between the energy shifts, but this would hardly be robust against the length of wire and perturbations of $J_0$.

Finally, we present an unified scenario for the shortest transfer time achievable via PP-transfer for $n_s=1,2,3$ excitations in the sender block in Fig.~\ref{fig:tim3rm}. We have also added the case $n_s=4$ to highlight its qualitatively similar behaviour to PP-transfer. Here we assume that the sender and receiver are connected by a wire able to transfer from one to four excitations by weakly coupling the respective blocks to the end to the wire. In order to have a wire able to perform such a task, its length $n_w$ has to fall in all the equivalence classes allowing PP transfer for $n_s=1,2,3$ and quasi-PP for $n_s=4$. Whereas $n_w$ can be arbitrary for $n_s=1,2$, for $n_s=3,4$, the length of the wire has to be $n_w=4l+1$ and $n_w=5l+1$ or $n_w=5l+2$, respectively. This yields to wires length of $n_w=20l+1$ and $n_w=20l+17$, respectively. In Fig.~\ref{fig:tim3rm}, we report the transfer times for the former case, noticing its linear increase with the wire's length for $n_s=2,3,4$. On the other hand, for $n_s=1$, the increase is $\sqrt{n_w}$ as the frequency involved in the PP transfer is derived from resolving the degeneracy via first-order perturbation theory, as shown for odd $N$ in Sec.~\ref{Ss.PPtransfer}.


\section{Applications of many-body PP transfer}\label{S.Applications}
In the previous sections, we have shown that  perturbatively perfect transfer of $n$ excitations is possible between the edges of a quantum wire. Now we analyse some cases where PP transfer is applied to the transport of relevant physical quantities, such as magnetisation and energy, highlighting first the invariance with respect to the fermionic or bosonic nature of the excitations.

\subsection{Equivalence of fermionic and bosonic observable's dynamics}\label{Ss.Observable}
An arbitrary one-body observable in second quantisation is given by
\begin{align}
\label{E_obs}
	\hat{O}=\sum_{nm}a_{nm}\hat{c}^{\dagger}_n\hat{c}_m+h.c.~,
\end{align}
where the $\hat{c}$'s are bosonic or fermionic operators acting on site $n$ and $m$. Expressing average of the observable's dynamics in Heisenberg representation, where $\hat{H}$ is given by Eq.~\ref{E_Ham1} yields
\begin{align}
\label{E_obs_1}
\average{\hat{O}(t)}=\sum_{nm}a_{nm}\sum_{kq}\phi_{kn}\phi^*_{qm}e^{i\left(E_k-E_q\right)t}\average{\hat{c}^{\dagger}_k\hat{c}_q}~.
\end{align}
As the single-particle spectrum of the Hamiltonian in Eq.~\ref{E_Ham1} is identical both for fermions and bosons, the only difference between the dynamics of an observable on a fermionic or bosonic many-body system that can possibly arise has to come from the average on the initial state of the operators on the RHS of Eq.~\ref{E_obs_1}.

In our setting the initial state is given by one or zero excitations per site
\begin{align}
\label{E_initial}
\ket{\Psi(0)}=\prod_{i=\{n_s\}}\hat{c}_i^{\dagger}\ket{0}~,
\end{align}
which is also the only initial state that fermions and bosons can have in common. Evaluating the average on the RHS of Eq.~\ref{E_obs_1} on this initial state reads
\begin{align}
\label{E_1Wick}
\bra{0}\hat{c}_1\hat{c}_2\dots \hat{c}_{n_s}\hat{c}^{\dagger}_k\hat{c}_q\hat{c}_{n_s}^{\dagger}\dots\hat{c}_2^{\dagger}\hat{c}_1^{\dagger}\ket{0}~.
\end{align}
Expressing all operators in the position basis reads
\begin{align}
\label{E.Wick1a}
\sum_{ij} \phi_{ki} \phi_{qj}^* \bra{0}\hat{c}_1\hat{c}_2\dots \hat{c}_{n_s}\hat{c}^{\dagger}_i\hat{c}_j\hat{c}_{n_s}^{\dagger}\dots\hat{c}_2^{\dagger}\hat{c}_1^{\dagger}\ket{0}~.
\end{align}
By straightforward application of Wick's theorem, we notice that the non-zero fully-contracted terms are those having an even number of permutations. As a consequence, the dynamics of an arbitrary one-body observable, such as in Eq.~\ref{E_obs_1}, is independent of the bosonic or fermionic nature of the excitations. For instance, the average number of particles on a lattice site, $\average{\hat{n}(t)}=\average{\hat{c}_n^{\dagger}(t)\hat{c}_n(t)}$ is the same whether the Hamiltonian in Eq.~\ref{E_Ham1} refers to bosons or fermions, notwithstanding Pauli's exclusion principle holds for fermions whereas bosons allow for multiple occupation.

It is easy to show that the same holds for $n$-body observables of the form
\begin{align}
\label{E_obs_n}
\hat{O}=\sum_{nmijrs\dots}\alpha_{nm...}\hat{c}^{\dagger}_n\hat{c}_m\hat{c}^{\dagger}_i\hat{c}_j\hat{c}^{\dagger}_r\hat{c}_s\dots+h.c.~,
\end{align}
when the average is evaluated on an initial state of the form of Eq.~\ref{E_initial} and the dynamics is ruled by a quadratic Hamiltonian such as in Eq.~\ref{E_Ham1}. A relevant example of a 2-body observables of the form of Eq.~\ref{E_obs_n} independent from the statistics of the excitations is the density-density fluctuations $\average{\hat{n}_i(t)\hat{n}_j(t)}$.

\subsection{Magnetisation transport}
\begin{figure}
	\centering
	\includegraphics[width=\linewidth]{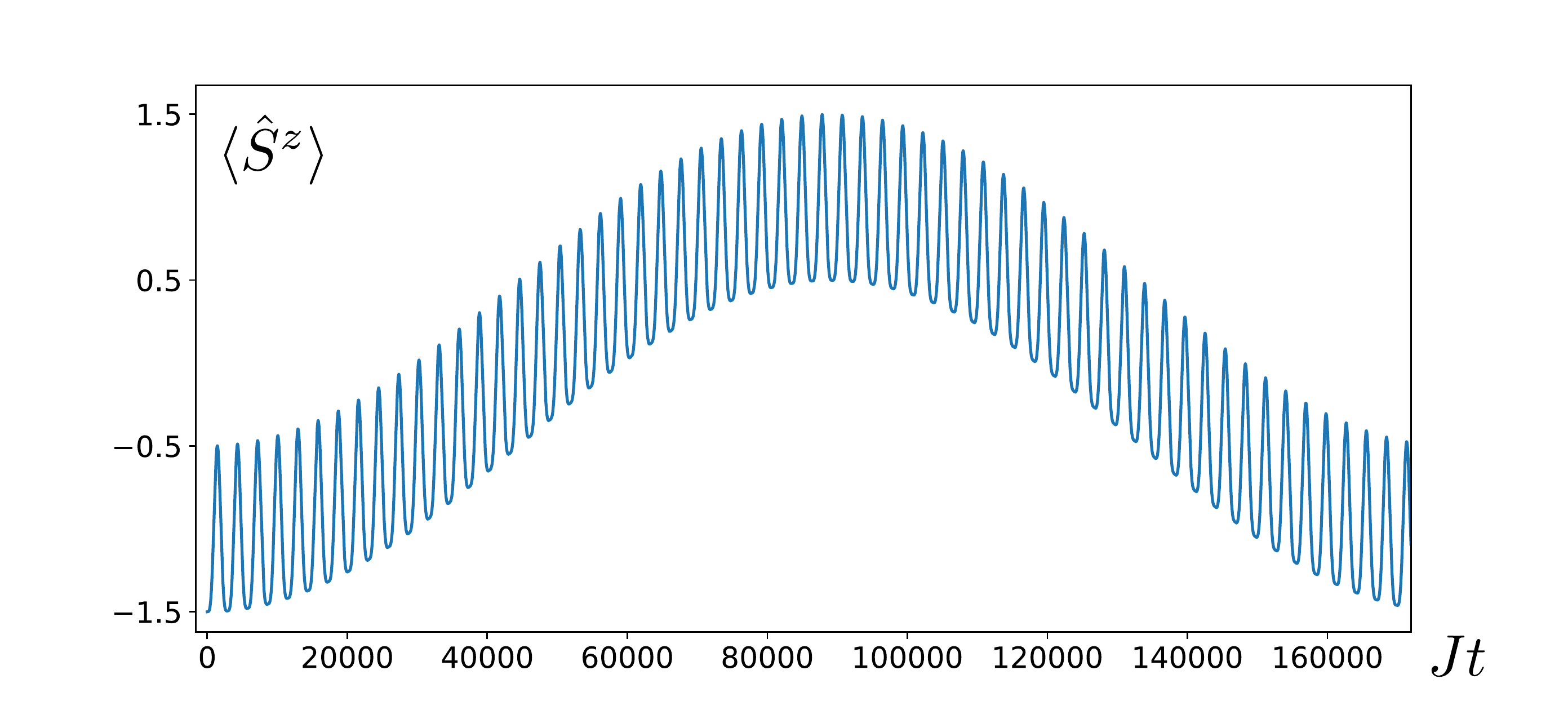}
	\caption{Average magnetisation of the receiver block, Eq.~\ref{E.Magn}, for $n_s=3$ in a chain of $N=47$. Notice that the occupation number entering Eq.~\ref{E.Magne} oscillates between two and three on a timescale much larger than the transition probability reported in Fig.~\ref{fig:dyn3}}
	\label{fig:we3}
\end{figure}

As it is well known, the Hamiltonian in Eq.~\ref{E_Ham1} models also a 1D spin-$\frac{1}{2}$ chain with isotropic interactions on the $XY$ plane, i.e.,
\begin{align}\label{Eq:22}
\hat{H} = \sum_{i}^{N}J_i\left(\hat{S}^{x}_{i}\hat{S}^{x}_{i+1} + \hat{S}^{y}_{i}\hat{S}^{y}_{i+1}\right)+h_i\hat{S}_i^z
\end{align}
when the standard Jordan-Wigner transformation is carried out~\cite{LIEB1961407}.
Because of the Jordan-Wigner mapping, the (fermionic) Hamiltonian in Eq.~\ref{E_Ham1} can model an $XX$ spin-$\frac{1}{2}$ open chain, where the average total magnetisation (along the $z$-direction) of a set of spins residing on sites $\{i\}$ is given by $\average{\hat{S}^z}=\sum_{\{i\}}\average{\hat{S}^z_i}=\sum_{\{i\}}\frac{2\average{\hat{c}_i^{\dagger}\hat{c}_i}-1}{2}$. As a consequence, the magnetisation of the receiver block evolves as
 \begin{align}
 \label{E.Magne}
 \average{\hat{S}_{\{r\}}^z(t)}=\sum_{\{r\}}\average{\hat{c}_r^{\dagger}(t)\hat{c}_r(t)}-\frac{n_r}{2}~,
  \end{align}
  where the average is evaluated over the initial state having all the spins in the sender block flipped. Straightforward calculations, in the Heisenberg picture and using Wick's theorem, allow to express the receiver's block magnetisation as a function of single-particle transition amplitudes $f_i^j(t)$: \begin{equation}\label{E.Magn}
\average{\hat{S}_{\{r\}}^z(t)}=\sum\limits_{i=\{s\}}\sum\limits_{j=\{r\}}|f_{i}^{j}|^{2}-\frac{n_r}{2}\equiv ||\mathcal{F}_{s}^{r}(t)||^2_{F}-\frac{n_r}{2}~,
\end{equation}
where $||\bullet||_{F}$ is the Frobenius matrix norm and $\mathcal{F}_{s}^{r}(t)$ is the submatrix defined in Eq.~\ref{E.FsubMatrix2}.
The result for $n_s=3$ is shown in Fig.~\ref{fig:we3}. Notice that, although the transition probability oscillates, for bosons, between 0 and 1 on a timescale of $J$ in the the corresponding scenario in Fig.~\ref{fig:dyn3}, the average number of bosons on the receiver block varies only between 2 and 3. Therefore, on a large time interval, with respect to $J$, around the transfer time $\tau$ at least two excitations out of three are located on the receiver block irrespective of their bosonic of fermionic nature. 
Indeed, the dynamics of the occupation number $\average{\hat{n}_i(t)}$ of site $i$ entering Eq.~\ref{E.Magne} is identical for bosons and fermions, as by the argument of Sec.~\ref{Ss.Observable}. As an example, let us consider the case $n_s=2$. Although the dynamics of the transition probability differs, as reported in Fig.~\ref{F.2ex_bos_fer}, the subspace spanned by the two photons in the receiver block is composed by $\{\ket{11},\frac{1}{\sqrt{2}}\left(\ket{02}\pm\ket{20}\right)\}$, which are all states having the same $\average{\hat{n}_i(t)}$.

\subsection{Energy transport}
The transfer of energy from one spatial location to another has always been a central topic in physics. Recently, a lot of attention has been devoted to the so-called quantum batteries, i.e., quantum devices able to store energy and release it upon demand at specific times~\cite{Alicki2013, Binder_2015, 2018arXiv180505507C, 2019arXiv191002458G}.
Devising a protocol to extract the maximum amount of energy from a charged battery, establishing a bound on its amount, and stabilizing the battery's charge has been addressed in several works~\cite{AllahverdyanRNEPL04,PhysRevLett.111.240401, 2018arXiv181104005J, PhysRevB.100.115142, PhysRevE.100.032107}
Another line of research is embodied by the investigation of the charging protocol a quantum battery~\cite{PhysRevLett.118.150601, PhysRevLett.122.047702, PhysRevB.98.205423}, and, apart from a few instances~\cite{PhysRevA.97.022106}, mainly non-interacting systems  embodying the quantum battery have been considered.

Our work can be immediately rephrased in terms of a charging protocol of a many-body quantum battery. Dubbing the sender block as charger, the receiver block as battery and the wire as a quantum cable connecting the charger to the battery, a quite natural set-up for charging a quantum battery is represented by Fig.~\ref{F.Figure}.

Nevertheless, in order to reinterpret the excitations dynamics in Sec.~\ref{S.Model} as a charging protocol, a few precautions are in place. As shown in Ref.~\cite{PhysRevB.98.205423}, the charging protocol should involve a time-dependent Hamiltonian 
\begin{align}
\label{E.Ham_bat}
\hat{H}(t) = \hat{H}_{0} +   \lambda (t)\hat{H}_{1}
\end{align}
where $\hat{H}_{0} = \hat{H}_{C} +\hat{H}_{w}+ \hat{H}_{B}$, are the time-independent Hamiltonians of the charger, the wire, and the battery, respectively. $\hat{H}_{1}$ is the Hamiltonian connecting the charger (battery) to the wire and $\lambda(t)$ is the coupling constant responsible for switching on and off the interaction between the charger (battery) and the wire when the charging protocol starts and ends. Generally, it is assumed that $\lambda(t)$ is  given by a step-function having a value of 1 for $t\in [0,\tau ]$ and $0$ otherwise. 
Because of that time dependence, energy may not be conserved and there could be some switching energy $\delta\textrm{E}_{sw}$ injected or extracted from the system during the protocol. This can be evaluated~\cite{PhysRevB.98.205423} by 
\begin{equation}\label{Eq:3q}
\delta\textrm{E}_{sw}(\tau ) = \textrm{tr}\left[\hat{H}_{1}\left(\rho(0) - \rho(\tau)\right)\right]~,
\end{equation} 
where $\rho$ is the density matrix of the full system.
Clearly, for $\comm{\hat{H}_0}{\hat{H}_1}=0$, one has $\delta\textrm{E}_{sw}(\tau )=0$. However, this is not the case for our model, since the commutator does not vanish. Nevertheless, evaluating Eq.~\ref{Eq:3q}, we obtain a zero switching energy due to the mirror-symmetry of our model. 
The two terms entering Eq.~\ref{Eq:3q} are equivalent to
\begin{align}
&\textrm{tr}\left[\hat{H}_{1}\rho(0)\right]=\bra{\Psi(0)}\hat{c}^\dagger_{n_B}\hat{c}_{w_1}+\hat{c}^\dagger_{n_w}\hat{c}_{1_B}+h.c.\ket{\Psi(0)}~,\label{E.E_sw1}\\
&\textrm{tr}\left[\hat{H}_{1}\rho(\tau)\right]=\bra{\Psi(0)}\hat{c}^\dagger_{n_C}(\tau)\hat{c}_{w_1}(\tau)+\hat{c}^\dagger_{n_w}(\tau)\hat{c}_{1_B}(\tau)+h.c.\ket{\Psi(0)}~,\label{E.E_sw2}
\end{align}
where the last equation is written in the Heisenberg representation. Eq.~\ref{E.E_sw1} is identically null because of the choice of the initial state of our system, whereas, expressing Eq.~\ref{E.E_sw2} in terms of single-particle transition amplitudes, results in
\begin{align}
\label{E.en_sw_f}
\textrm{tr}\left[\hat{H}_{1}\rho(\tau)\right]=  \sum_{\substack{n\in\{n_C\}}}
\left((f_{n}^{n_C})^*f_{n}^{1_w} +\left(f_{n}^{n_w}\right)^*f_{n}^{1_B}+ c.c.\right)~,
\end{align}
with $c.c.$ denoting complex conjugation.
The above expression is identically null as each $(f_{n}^{i})^{*}f_{n}^{i+1}$ results to be purely imaginary according to the conditions outlined in Sec.~\ref{Ss.MBD} for mirror-symmetric matrices.

As a consequence, the figures of merit for the charging protocol of a quantum many-body system via a quantum wire, are those reported in Ref.~\cite{PhysRevB.98.205423}. The mean energy stored in the battery and the mean storing power are, respectively,
\begin{equation}\label{Eq:4q}
\textrm{E}_{B}(\tau ) = \textrm{tr}[\mathcal{H}_{B}\rho_{B}(\tau )]~,~P_{s}(\tau ) =\frac{\textrm{E}_B(\tau )}{\tau}~.
\end{equation} 

Other useful quantities are the maximum energy stored and the maximum power,
\begin{equation}\label{Eq:6q}
\bar{\textrm{E}}_{s}(\tau ) \equiv  \max_{\tau}[\textrm{E}_{s}(\tau )]\equiv\textrm{E}(\bar{\tau})~,~
\tilde{P}_{s}(\tau ) \equiv  \max_{\tau}[P_{s}(\tau )]
\end{equation}
and their corresponding optimal charging times,
\begin{equation}\label{Eq:8q}
\bar{\tau}\equiv \min_{\textrm{E}(\bar{\tau})=\bar{\textrm{E}}_{s}(\tau )}[\tau ]~,~ \tilde{\tau}\equiv \min_{P(\tilde{\tau})=\tilde{P}_{s}(\tau )}[\tau ]~.
\end{equation}
Lastly, the charging power obtained at maximum energy is defined as,
\begin{equation}\label{Eq:9q}
\bar{P}_{s}(\tau ) \equiv \frac{\bar{\textrm{E}}_{s}(\tau )}{\tau}~,
\end{equation}
which is generally different from the maximum power because the times at which maximum energy and maximum power are achieved, $\bar{\tau}$ and $\tilde{\tau}$ respectively, may not coincide. 

In this subsection we choose $h>1$, so that the charger state with all spin aligned in the positive $z$-direction is the highest energy eigenstate of $\hat{H}_B$, with energy $\frac{n_B h}{2}$. Applying the same magnetic field $h$ to the rest of the system, wire and battery, allows us to use the formalism of Sec.~\ref{S.Model} to evaluate the above figures of merit, as a uniform magnetic field in Eq.~\ref{E_Ham1} implies only an uniform shift by $h$ of all single-particle eigenenergies, with the eigenvectors remaining unchanged. Clearly such a uniform shift brings along only an irrelevant overall phase factor in the dynamics.

Interestingly, only the one-body terms in $\hat{H}_B$ contribute to the mean energy $\textrm{E}_{B}(\tau )$ in Eq.~\ref{Eq:4q}. This can be immediately seen as, at time $\tau$, the density matrix of the battery $\rho_{B}(\tau)$ represents the state with all the spins flipped as PP-transfer has occurred. In addition, it results also that the energy due to the inter-spin interaction term is vanishing at all times, as a result of the following equation
\begin{align}
\label{E.interaction_en}
\textrm{E}_{I}\equiv \sum_{\substack{i\in \{n_B\}}}
\frac{1}{2}\average{\hat{c}_{i+1}^{\dagger}\hat{c}_{i}+h.c.} = \sum_{\substack{i\in \{n_B\}\\n\in\{n_C\}}}
\left((f_{n}^{i})^{*}f_{n}^{i+1} + c.c.\right)~,
\end{align}
again because of the conditions on $f_i^j(t)$  for mirror-symmetric matrices, as already derived  for the switching energy $\delta E_{sw}$.

\begin{figure}
	\label{F.Energy}
	\includegraphics[width =0.535\textwidth]{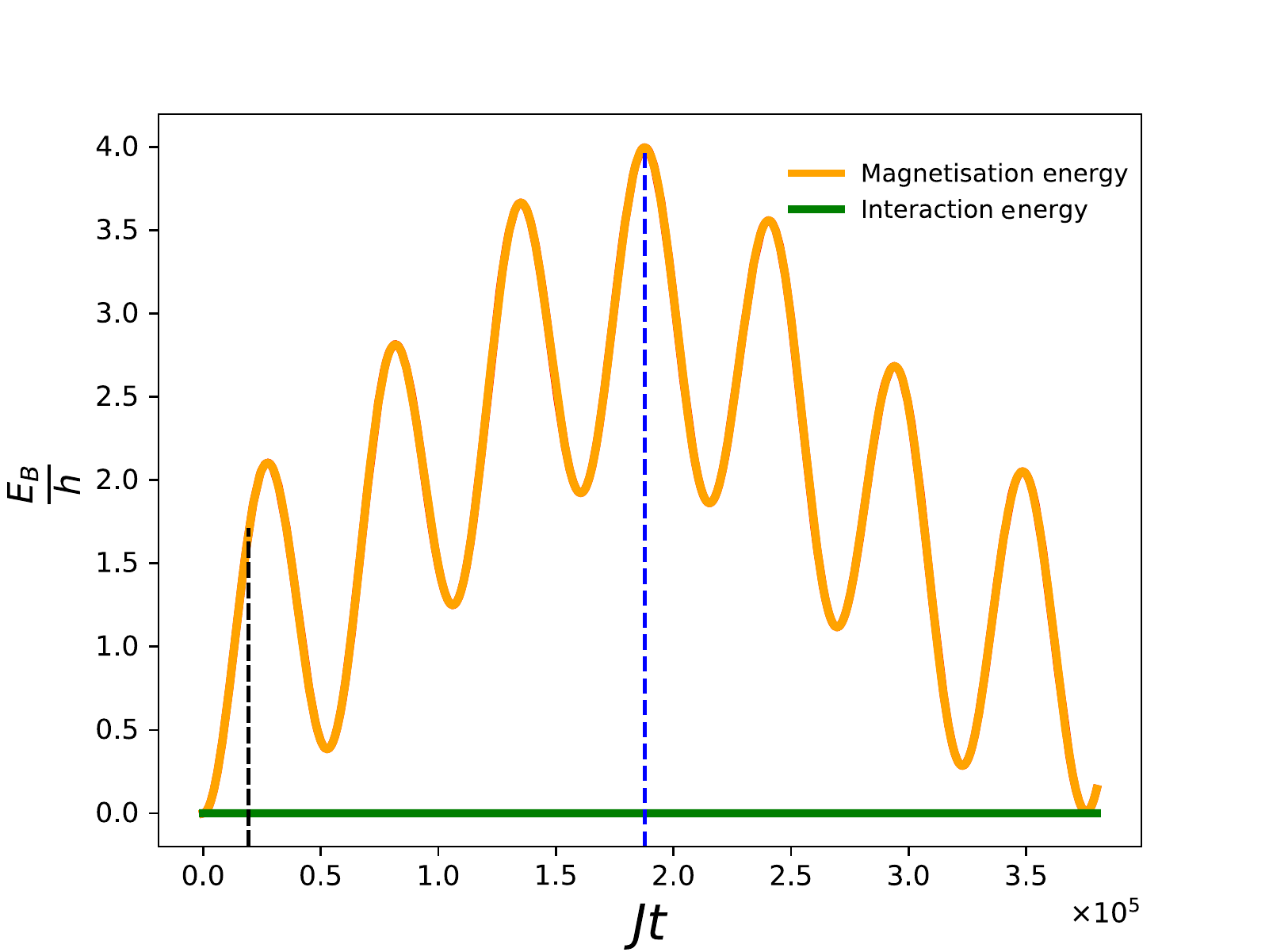}
		\includegraphics[width =0.47\textwidth]{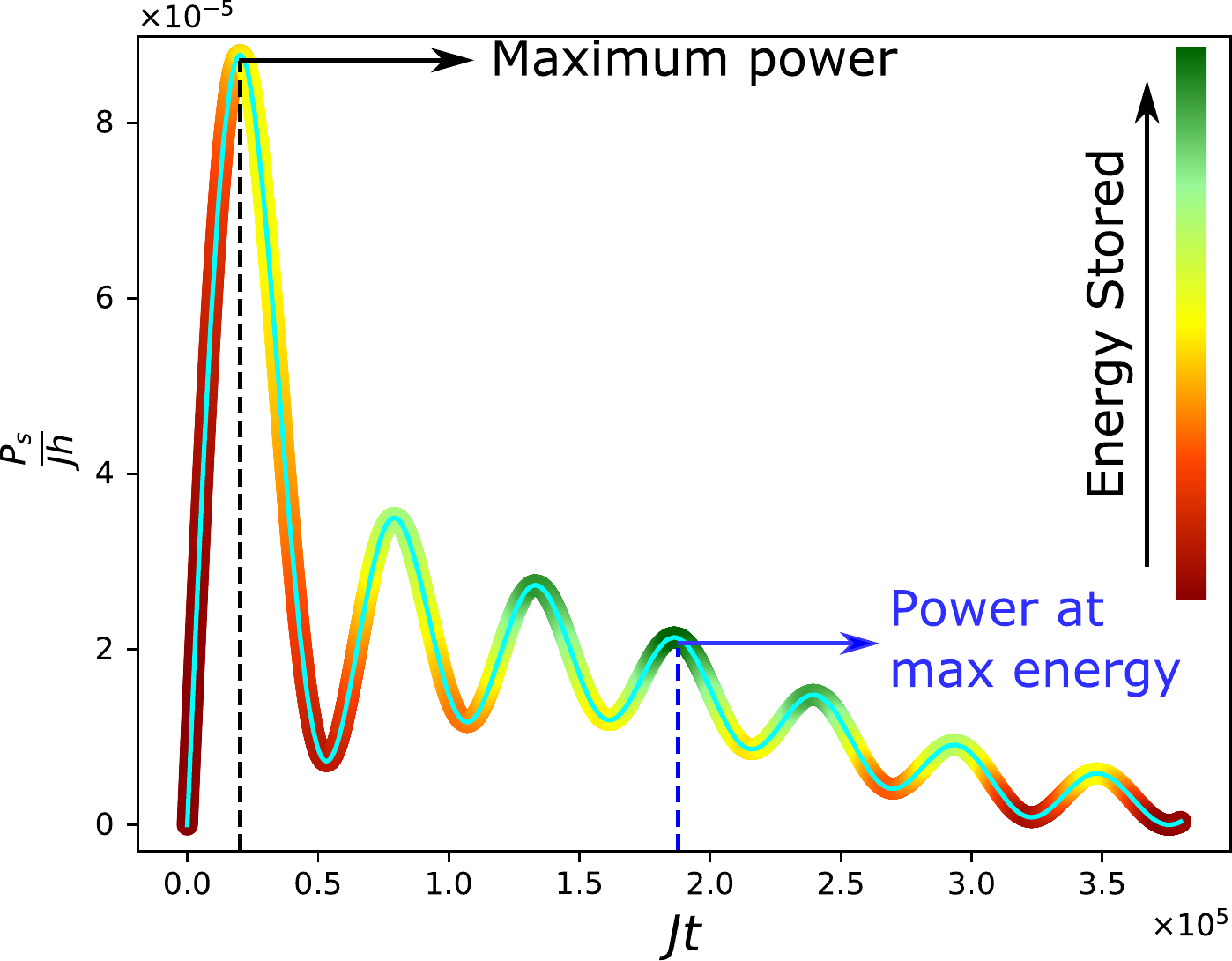}
	\caption{Energy and charging power for a battery made out of 4 qubits $n_B=4$, left and right panel, respectively. The dotted line in the left panel corresponds to the transfer of the four excitations from the sender to the receiver. The orange line is the contribution of the one-body term in $\hat{H}_B$, whereas the green line that of the two-body term. In the right panel, the black dotted line is the time at which the maximum power is achieved, whereas the blue dotted line is the time at which the maximum energy is stored in the battery, Eq.~\ref{Eq:9q}. The corresponding energy stored in the battery is reported in the left figure by the same lines.  The color of the curve $P_{s}(\tau )$ indicates the amount of energy stored in the battery at that time. The length of the wire is 32, for a total length of the chain of 40.}
\end{figure}

This allows us immediately to use our results to confirm that all the charger's energy is transferred to the battery and, remarkably, no energy is stored in two-body correlations at any time. This has several advantages: on the one hand, only single-qubit operation are necessary to extract the energy from the many-body battery and, on the other hand, the $n_B$ spins embodying the battery can be split in independent, non-interacting partitions without any loss of the initially stored energy.
An instance of the charging process of a quantum many-body battery is shown in Fig.~\ref{F.Energy} for the case of $n_s=4$. Notice that the power at maximum energy as by Eq.~\ref{Eq:9q} is obtained at a considerably larger time than the maximum power, Eq.~\ref{Eq:6q}. 

\section{Conclusion}\label{S.Concl}
We have investigated the many-body transfer of bosonic and fermionic excitations in a one-dimensional open chain modeled by a nearest-neighbor hopping Hamiltonian. The set-up consisted of a block of sender sites, each hosting one excitation, weakly coupled to a quantum wire at one edge with the block of receiver sites weakly coupled at the opposite edge. We have found that up to three excitations can be transferred between the edges of the chain in a regime dubbed PP-transfer, perturbatively perfect, via Rabi-like dynamics. We have identified the lengths of the wire that allow PP-transfer analyzing the occurrence of first- and second-order perturbative corrections to the energy degeneracies among the single-particle energy levels of the sender (receiver) block and the wire. This has yield us to identify modular arithmetic equivalence classes of the wire's lengths supporting PP transfer. Consequently, we have found that, for a number of excitations greater than two, not all length of the wire exhibits PP transfer, at variance for the case of one and two excitations. 

The results obtained have then been applied to the investigation of the dynamics of two physically relevant quantities: magnetisation and energy transport. In the former case, we obtained that the receiver spins get fully magnetised at the PP-transfer time and, moreover, partial magnetisation is retained for a long time. The energy transfer protocol has been investigated in the framework of quantum battery charging, one of the few instances where the quantum battery consists of a many-body system, and we obtained a complete charging of the battery with energy stored only in the on-site interaction Hamiltonian term. This has the advantage that an energy extraction protocol needs to consists only of local operations on each site. We also showed that relevant physical quantities, such as the average number of excitations in the sender block, is the same both for a fermionic and a bosonic chain.

Due to the quadratic nature of the Hamiltonian, we were able to investigate the excitations dynamics for arbitrary lengths of wire, reducing every quantity under scrutiny to functions of single-particle transition amplitude. It would be interesting to investigate whether PP transfer occurs also in interacting models, and, if so, if there are the conditions on the wire's length. Another interesting application of our results could be in quantum information processing. In Refs.~\cite{PhysRevLett.106.140501,PhysRevA.100.052308} it has been shown that, at half the transfer time, the sender and receiver block are maximally entangled for one- and two-particles in the sender block. Similarly, for an higher number of excitations in the sender block, a similar scenario occurs. Also investigating the quantum state transfer of an arbitrary state of $n$ qubits would have several applications in quantum information processing. Whereas weak-couplings has been shown to be successful for one- and two-qubits quantum state transfer~\cite{PhysRevA.72.034303, doi:10.1142/S021974991750037X, qst2}, the case of an higher number of qubits has not been yet addressed and will be our subject of further investigation. 

\section*{Ackowledgements}
The authors thank Prof.~Andr\'{e} Xuereb and Dr.~Zsolt Bern\'{a}d for useful discussions. C. S. acknowledges funding by the European Union's Horizon 2020 research and innovation programme under Grant Agreement No.\ 732894 (FET Proactive HOT).

\section*{References}

\bibliographystyle{unsrt.bst}

\bibliography{biblio.bib}

\end{document}